# Switching Current Measurements of $EuBa_2Cu_3O_y$ Intrinsic Josephson Junctions

**S. Ueda, T. Okutsu, S. Ishii and Y. Takano**
*National Institute for Materials Science, 1-2-1 Sengen, Tsukuba 305-0047, Japan*

We investigated the switching dynamics of two kinds of $EuBa_2Cu_3O_y$ (Eu123) intrinsic Josephson junctions (IJJs) with different anisotropy parameters $\gamma$ (= $(m^*_c/m^*_{ab})^{1/2}$) tuned to 37 and 22. In contrast to weakly-coupled IJJs in Bi2212, significant deviations from the thermally activated escape model of a single-stack Josephson junction were observed in their switching current distributions, $P(I)$, due to the strongly-coupled nature of the Eu123 IJJs, The $P(I)$ of the two IJJs are found to be independent of temperature, below 1.4 K and 4.2 K, respectively; which indicates the observation of macroscopic quantum tunneling at high temperatures of liquid He.

In the last few years, the observation of macroscopic quantum tunneling (MQT) and energy level quantization have been successfully demonstrated in YBCO grain boundary junctions[1,2] and in Bi2212 IJJs.[3-6] This result supports the idea that high $T_C$ superconductor (HTS) materials can be used as so-called phase qubits based on MQT, which can be used for quantum information processing. In particular, in Bi2212 IJJs, where the influence of dissipative quasi-particles that suppresses MQT is very small,[7-10] the MQT has been observed at relatively higher temperatures than that in conventional Josephson junctions of low $T_c$ superconductors (LTS). This is attributed to the high plasma frequency, $\omega_{p0}$, of HTS with a high critical current density, $J_c$, which is highly advantageous for quantum device applications. Since IJJs are consisted of alternating superconducting and barrier layers that are much thinner than the magnetic penetration length, they show a strong interaction along the direction perpendicular to the layers. Recently, an important result of IJJs was reported by Jin *et al.*[4] ; the MQT escape rate $\Gamma$ is enhanced proportionally to the square number of junctions, $N^2$ , in Bi2212 due to the "globally coupling" IJJs.

In the present study, we examined the switching dynamics of IJJs in Eu123. The IJJs in a rare earth 123 (RE123) system exhibit a particularly low $\gamma$ compared to known HTS, which leads to a high $J_c$ perpendicular to the *c*-axis. Therefore, they have a significantly higher $\omega_{p0}$ (up to a few THz)[11,12] and are more strongly- coupled junctions than those of Bi2212, which provides a higher crossover temperature in the MQT process compared to Bi2212. In this study Eu123 was heavily under-doped where the $\gamma$ was 22 – 40 (the barrier layers of RE123 are not insulating around the optimal doped region), enhancing the electrical isolation of the barrier layers. $J_c$ in this case was in the range of 10 – 100 kA/cm$^2$ at 4.2 K.

Eu123 single crystal whiskers were grown using a Sb-doping method,[17] and two kinds of inline junctions, sample A and B, were fabricated from whiskers using a focused $Ga^+$ ion beam. The details of the fabrication process are described elsewhere.[11,18] Post-annealing was then performed only for sample A at 400°C under vacuum ($P_{O2}$ < $10^{-2}$ Pa). Fig. 1 is an SIM image of sample B. The overlap lengths through the *c*-axis of both the samples were 0.1 μm, which correspond to the total junction number of ~ 80. The capacitance, *C*, of the junctions were estimated to be 142 and 474 fF for sample A and B, respectively, using a dielectric constant of YBCO (= 15).[19] The parameters of the samples were estimated from measurements at 4.2 K, and are summarized in Table I.

The *I-V* characteristics of both the samples were measured using the four-terminal method in a $^3$He cryostat from room temperature to 0.6 K. The samples were biased using a constant ramp rate that was varied from 0.015 - 0.24 A/s, using an analog function generator. The observed signals were multiplied using the low noise self-made amplifier and displayed and measured using an oscilloscope. The switching current was measured using the "single-pulse-method", where individual voltage pulses were discontinuously generated from the function generator in order to eliminate the self heating effect on the effective measurement, and at each temperature the switching current was recorded 1000 – 8000 times by detecting voltage jumps.

The $P(I)$ of the Eu123 IJJs were fitted using a single Josephson junction (SJJ) model, which is usually used for multi-stack IJJs. As is well known, the Kramers theory in the low damping regime gives the escape rate $\Gamma$ from the $V$ = 0 state of the SJJ, which is given by[13,14]:

$$\Gamma = a_t \frac{\omega_0}{2\pi} \exp\left(-\frac{U_0}{k_B T}\right) \qquad (1)$$

where $U_0$ is the barrier energy of the meta stable state. The meta stable state is found to decrease for an increasing bias current: that is, $U_0(I) = h/2e[2I\sin^{-1}(I/I_c)+2(I_c^2-I^2)^{1/2}-\pi I]$. Here $I$ and $I_c$ are the bias and the critical current, respectively, $\omega_0 = [1-(I/I_c)^2]^{1/4}(2\pi I_c/\Phi_0 C)$ is the bias depending plasma frequency, and $\Phi_0$ is the magnetic flux quantum.

Table I Junction size, $T_c$, $J_c$, $\gamma$, $C$, $\omega_{p0}$, and $\beta_c$ at 4.2 K for Eu123 samples.

| Sample | A | B |
|---|---|---|
| Junction size [μm$^3$] | 0.9 x 1.0 x 0.1 | 1.5 x 2.0 x 0.1 |
| $T_c$ [K] | 34.5 | 40.5 |
| $J_c$ [Acm$^{-2}$] | 1.47 x $10^4$ | 8.7 x $10^4$ |
| $\gamma$ (= $(m^*_c/m^*_{ab})^{1/2}$) | 37 | 22 |
| $C$ [fF] | 142 | 474 |
| $\omega_{p0}$ [GHz] | 290 | 657 |
| $\beta_c$ | 9 | 12 |

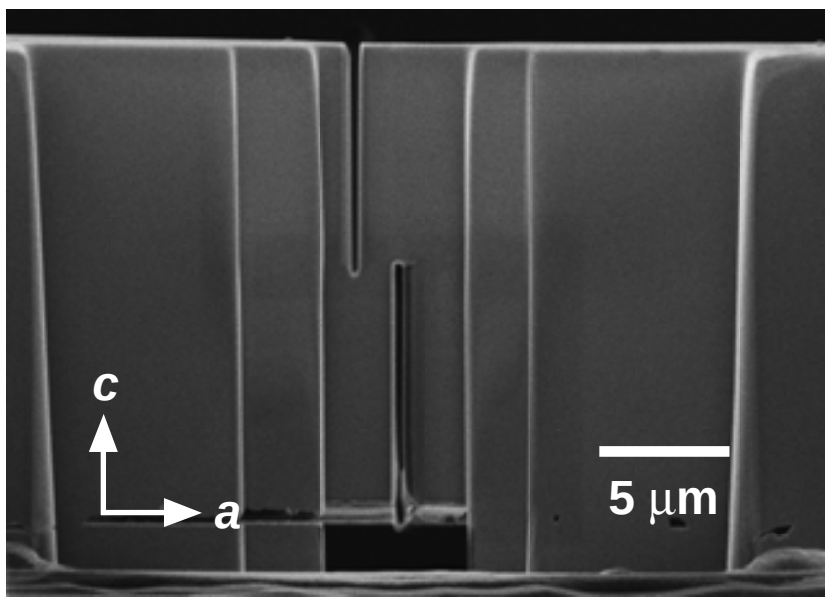

FIG. 1. SIM image of *ac*-plane of sample B.

$a_t$ is the temperature and the damping dependent thermal parameter. At temperatures below a crossover temperature, $T^*$, the escape is dominated by quantum tunneling through the barrier.

The *I*-*V* curves of sample A with $\gamma$ = 37 exhibited clear multiple branches; figure 2 (a) is a plot of the first switching from zero to the first branch at 15 K. In contrast, the *I*-*V* curves of sample B with $\gamma$ = 22, as shown in figure 2 (b), exhibited uniform switching in all junctions due to the stronger coupling in IJJs than that in sample A. In this case, there was no stable state between 0 mV ~ 400 mV, where all junctions were in the resistive state; this is the same behavior as "uniform-stack switching", recently observed in strongly-coupled Bi2212 IJJs.[4] The differences in the *I*-*V* characteristics in both the samples were mainly dependent on $\gamma$, which directly affects the $J_c$ and the Josephson coupling energy, $E_J$, of the samples. The McCumber parameter $\beta_c$ ($\approx (4I_c/\pi I_r)$) of the samples,[20] where $I_r$ is the return current, were estimated as 9 and 12 for sample A and B, respectively, from the *I*-*V* curves. From this it was found that both the samples were in the underdamped regime. The *I*-*V* curves of both the samples exhibited a clear hysteresis at all temperatures. Figure 3 (a) shows a plot of $P(I)$, for the first switching of sample A as a function of the bias current $I$, measured at 7.9 – 16.3 K. The open symbols represent the experimental data, and the solid lines represent the theoretical fitting based on equation (1).

Interestingly, at the high temperature of 16.3 K the experimental data fit well with the SJJ model, however, it was found to deviate gradually for a decreasing temperature, and broaden in the lower bias current region. The broadening of $P(I)$ corresponds to the enhancement of the escape rate in the lower bias current region, which can be attributed to the strong coupling in IJJs.

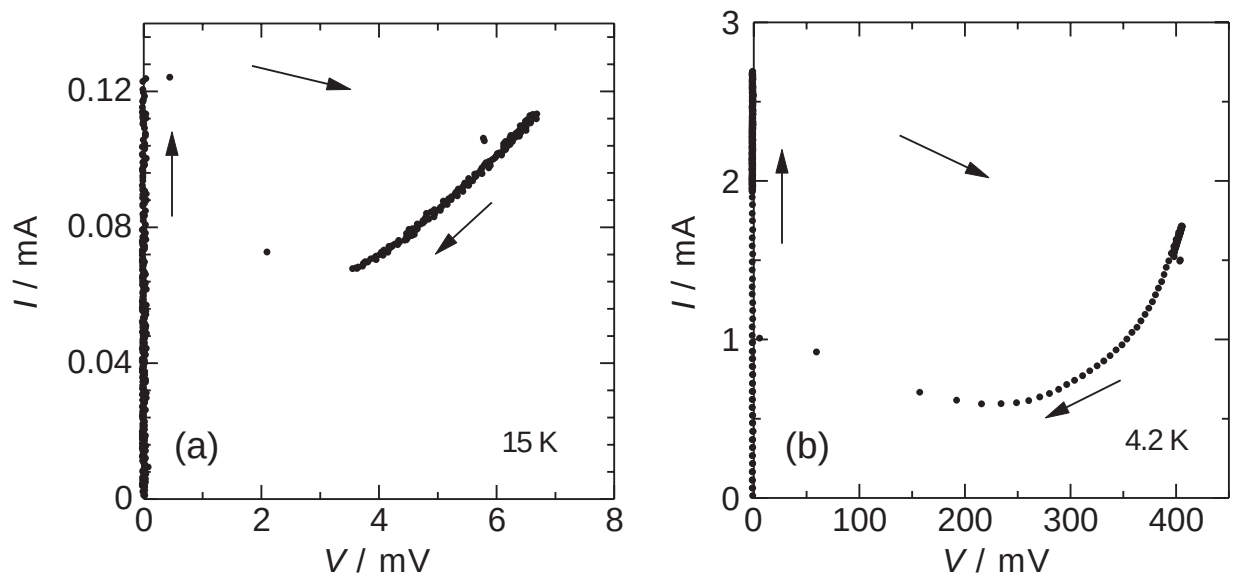

FIG. 2. *I*-*V* curves of (a) the first switching, corresponding to the transition from zero to the first branch of sample A, and (b) uniform switching in all junctions of sample B.

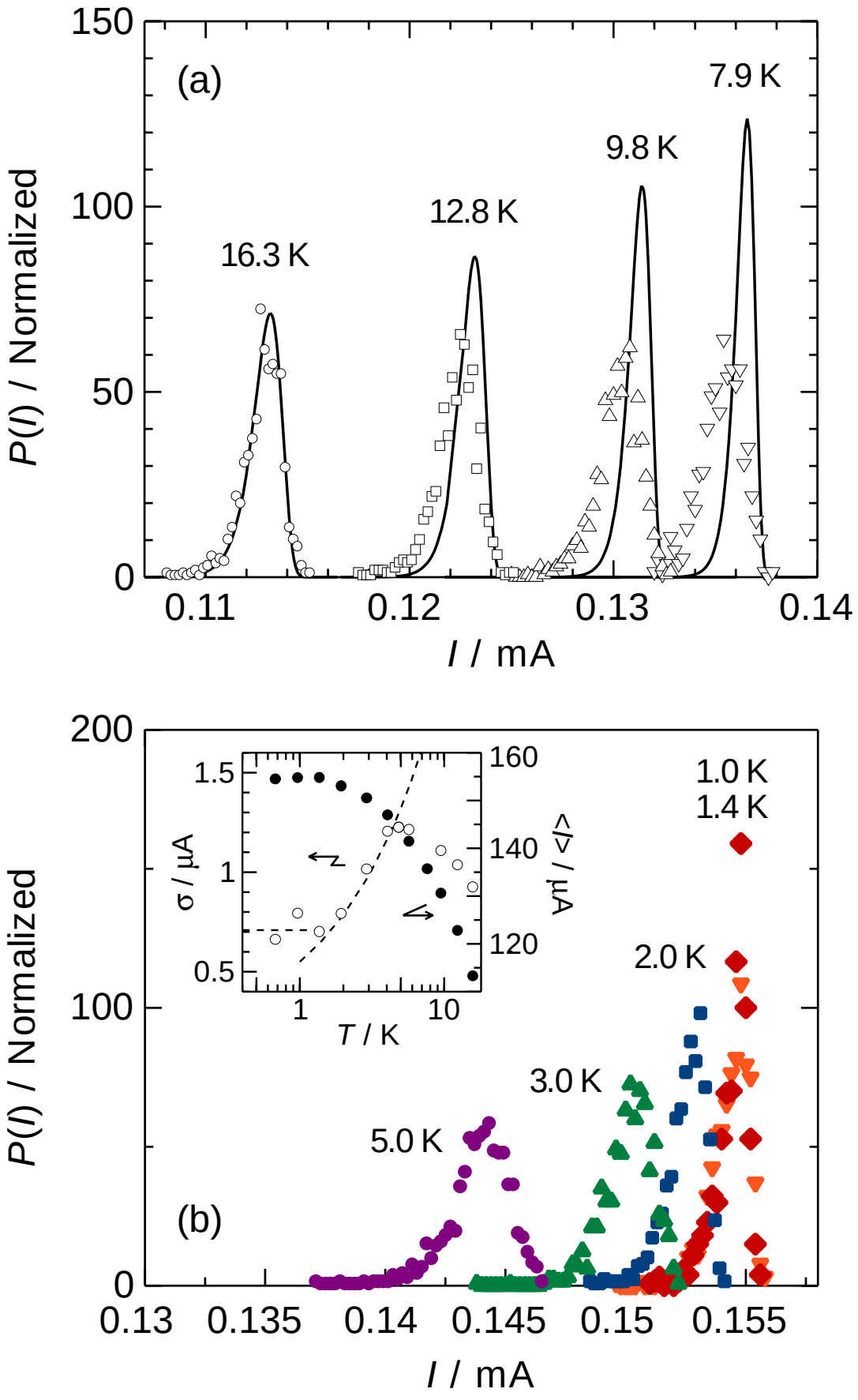

FIG. 3. (a) The experimental data (open symbols) and theoretical fitting (solid lines) of $P(I)$ for sample A measured at 7.9 – 16.3 K. (b) $P(I)$ of sample A in the lower temperature region 1.0 – 5.0 K, and the inset shows the distribution width $\sigma = (\langle I^2\rangle - \langle I\rangle^2)^{1/2}$ and $\langle I\rangle$ as a function of temperature.

In the case of Bi2212, the deviation from the SJJ model, leading to the increase of the escape rate, was observed only in the MQT process[4]; however, there was no deviation observed in the thermally assisted region. For the Eu123 IJJs it was found that the SJJ model could not be adopted even in the thermally assisted region because of the significantly lower $\gamma$ compared to Bi2212, and the strong coupling of the junctions. It is speculated that the effects of the IJJs coupling was prominent when the temperature was decreased because thermal fluctuation decreases.

Figure3 (b) is a plot of $P(I)$ from sample A in the lower temperature region of 1.0 – 5.0 K. Although the plot of $P(I)$ no longer agrees with the SJJ model, the peaks were found to shift gradually to higher currents and become sharp with decreasing temperature below 5 K, and independent of temperature at 1.4 K. The distribution width $\sigma = (\langle I^2\rangle - \langle I\rangle^2)^{1/2}$ and the mean value of the switching current $\langle I\rangle$ as a function of temperature are presented in the inset of figure 3 (b). The distribution width $\sigma = (\langle I^2\rangle - \langle I\rangle^2)^{1/2}$ and the mean value of the switching current $\langle I\rangle$ as a function of temperature are presented in the inset of figure 3 (b). It was also found that $\sigma$ decreases proportional to $T^{2/3}$ below 5 K, and $\sigma$ and $\langle I\rangle$ are saturated at 1.4 K, confirming that the escape process was dominated by MQT below 1.4 K. The $T^*$ of 1.4 K was higher than that of Bi2212.

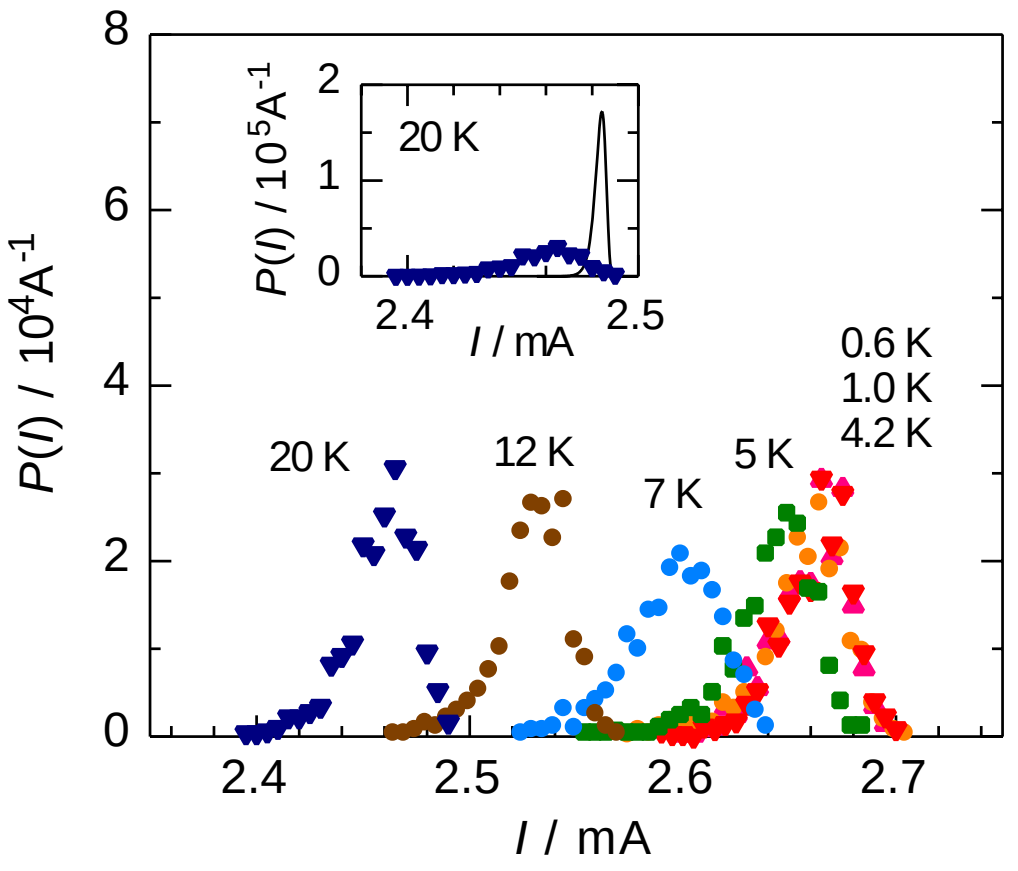

FIG. 4. $P(I)$ and $\Gamma(I)$ for sample B measured at 0.6 – 20 K, and the inset shows the comparison between the experimental data (dots) and theoretical fitting (solid lines) at 20 K.

There are two explanations for this: the first is due to the high frequency $\omega_{p0}$, and the second is the strong coupling nature of Eu123 IJJs.

Figure 4 is a plot of $P(I)$ as a function of the bias current $I$ for sample B, for 0.6 – 20 K. $P(I)$ was found to gradually shift to higher currents for a decreasing temperature and saturate at 4.2 K, indicating that the MQT occurs below 4.2 K. In this sample, the SJJ model could not be adopted for all temperate regions because the coupling was stronger than that of sample A, due to the lower $\gamma$. Since $P(I)$ was observed to broaden significantly for a lower current, the distribution width $\sigma$ is large, as in the inset of figure 4, and the systematic temperature dependency of $\sigma$ was not observed in this sample. In order to confirm MQT in such strongly-coupled IJJs, an extended escape model for multi-stack junctions is required instead of the SJJ model. $P(I)$ became independent of temperature at 1.4 K and 4.2 K for sample A and B, respectively; this is the highest $T^*$ that has been reported. Considering the higher $\omega_{p0}$ of the Eu123 IJJs, 290 and 657 GHz for sample A and B, (several times higher than that of Bi2212 (~100 GHz)), and the effect of the strong coupling of IJJs that enhances the MQT rate, the MQT can reasonably occur at temperatures several times higher than that of Bi2212.

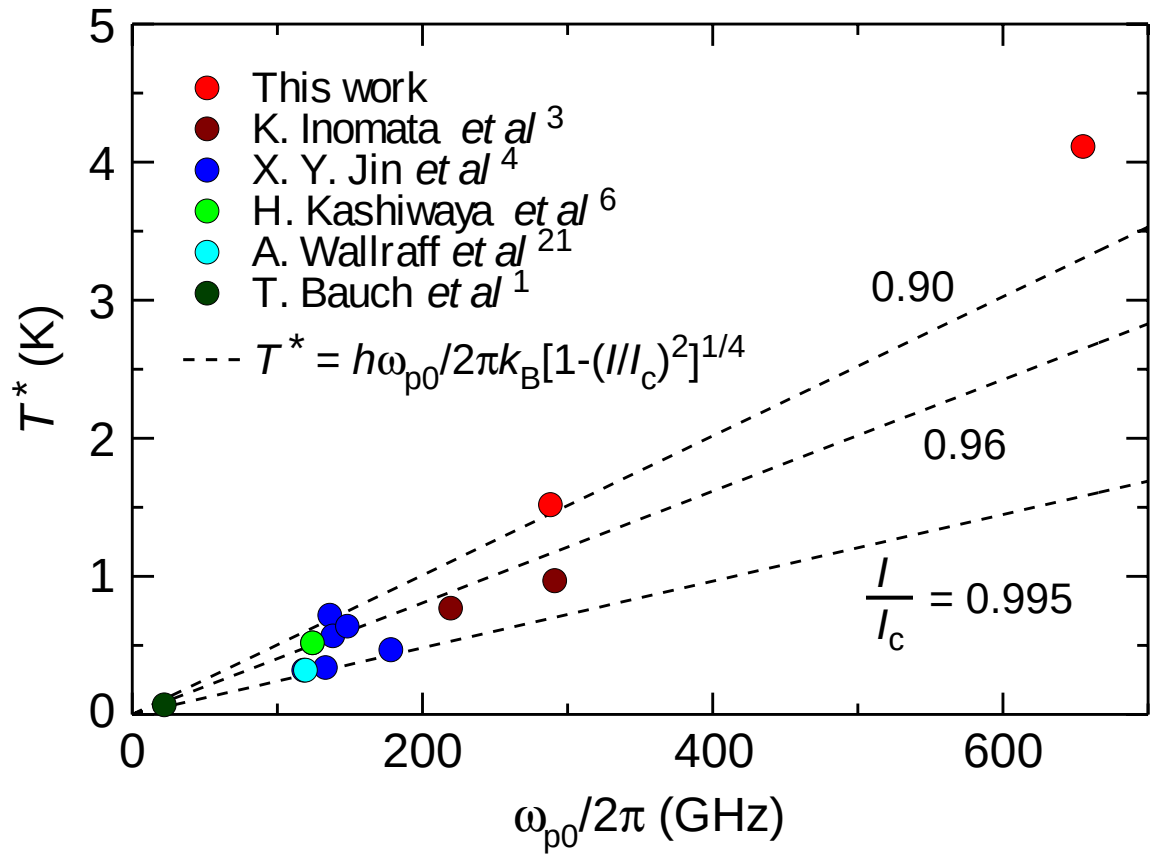

FIG. 5. The relation between $\omega_{p0}$ and $T^*$ for the sample A and B, the Bi2212 IJJs, and an IJJ of LTS [1,3-6,21]. The dotted lines represent the equation $T^* = \hbar\omega_{p0}/2\pi k_B[1-(I/I_c)^2]^{1/4}$.

In figure 5 is a plot of $T^*$ as a function of $\omega_{p0}$ for the samples used in this study, and data of the Bi2212 IJJs and a Josephson junction of LTS [1,3-6,21]. The dotted lines represent the equation $T^* = \hbar\omega_{p0}/2\pi k_B[1-(I/I_c)^2]^{1/4}$. The Bi2212 IJJs and a junction of LTS switch between $I/I_c$ ~ 0.96 – 0.995. In contrast, our data and the strongly-coupled Bi2212 IJJs reported by Jin *et al.,* are not plotted in the region, but in the lower $I/I_c$ region, resulting in a higher $T^*$ than usual, which is assumed to be the cause of premature switching due to the strong coupling of IJJs.

In conclusion, $P(I)$ of Eu123 IJJs exhibited significant broadening deviations from the escape model of a SJJ due to the strong coupling nature of IJJs in Eu123, corresponding to the enhancement of the escape rate. $T^*$ was found to be 1.4 K and 4.2 K for sample A and B, respectively. Form these results, we suggest that the IJJs in a RE123 system exhibit good potential for the application in phase qubits based on MQT at high temperatures.